\documentstyle[12pt,psfig,amssymb]{article}    

\begin{document}
\title{
\vspace{10mm}
Amplification of Isocurvature Perturbations induced
       by Active-Sterile Neutrino Oscillations
		}  

\author{P. Di Bari \\
{\small\em Dipartimento di Fisica, Universit\'a di Roma "La Sapienza" }\\
      {\small\em and I.N.F.N. Sezione di Roma I} \\
       {\small\em P.le A.Moro 2, I00185 Roma, Italy} \\
         {\small\em (dibari@roma1.infn.it)} }    
\date{}
\maketitle

\begin{abstract}
We show how the generation of a lepton number in the Early Universe 
induced by $\nu_{\alpha}\leftrightarrow\nu_{s}$,  
in presence of small baryon number inhomogeneities, 
gives rise to the formation of lepton domains. 
The structure of these domains reflects the spectral 
features of the baryon number inhomogeneities that generated it and
interestingly the size of the lepton domains can be super-horizon. 
\end{abstract}


keywords: {\small early universe, neutrino oscillations, sterile neutrino,
	    baryon isocurvature perturbations, leptogenesis.}

\section{Introduction} 

If one assumes the 
existence of a mixing between an active neutrino flavor 
$\alpha$ and a sterile neutrino flavor, then a lepton number can 
be generated in the early universe \cite{Foot96}. To this purpose a 
small vacuum mixing angle ($\sin 2\theta_{0}\ll 1$) is necessary 
(the mass eigenstates almost coincide with the interaction eigenstates). 
Moreover the sterile 
neutrino must be lighter than the active neutrino and the absolute 
value of $\delta m^2$, the difference of squared masses, must be 
in the range $(10^{-5} \div 10^5)\; {\rm eV^2}$. In fact the critical 
temperature at which the generation occurs is given, if the sterile 
neutrino production is until that time negligible, by the expression:
\begin{equation}\label{eq:TC} 
T_c\simeq 14.5\, (18.0)\, {\rm MeV} 
\left(\frac{|\delta m^2|}{\rm eV^2}\right)^{\frac{1}{6}}\hspace{10mm}
 \alpha=e\, (\mu,\tau)
\end{equation} 
It follows that if $|\delta m^2|\stackrel{<}{\sim} 10^{-5} 
{\rm eV^2}$, then $T_c\stackrel{<}{\sim} 3\, {\rm MeV}$ and the lepton 
number variation would be dominated by the MSW effect. In this case 
a lepton number larger than $10^{-7}$ cannot be generated 
\cite{Enqvist91,Barbieri91}. 

On the other hand if $|\delta m^2|\stackrel{>}{\sim} 10^{5} 
{\rm eV^2}$ then $T_c\stackrel{>}{\sim} 150{\rm MeV}$ and one should 
be able to describe the neutrino oscillations in a quark-gluon plasma, 
something beyond the present level of matter effects account (moreover it is 
not interesting in this context, because it would imply active neutrino 
masses much higher than $100\, {\rm eV}$, cosmologically excluded if 
neutrinos are stable). 

A rigorous description of neutrino oscillations in the early universe 
requires a quantum kinetic approach able to describe the evolution 
of the statistical density matrix for the two mixed neutrino flavors 
\cite{Dolgov81}.
Such a description must include the different effects of matter on the 
neutrino mixing: a coherent effect due to forward scattering \cite{Wolf78}, 
a loss of coherence due to the collisions that change the neutrino momentum
\cite{Harris82,Stodolsky87} and the repopulation of active neutrino states 
(depleted by the oscillations into the sterile neutrino states) through the 
collisions \cite{Thomson92,McKellar94}. 
 
 For a better understanding of the generation of a lepton number, 
it has been shown to be more convenient to turn to a simpler description 
that neglects the possibility of a MSW effect at the resonance. 
This approximation proves to be valid at temperatures 
$T\stackrel{>}{\sim}5\,{\rm MeV}$ because the collisions destroy
the coherence at the resonance.
In this way  the physics  underlying the lepton number generation is isolated and can 
be  clearly understood  \cite{Foot96,Foot97}. The role of collisions, from 
this point of view, is crucial: they can be usefully considered,
in a heuristic sense, as a measurement 
process able to make the two-quantum state collapsing in one of the two 
interaction eigenstates. Through this effect sterile neutrinos are produced
and the presence of a small baryon number induces a tiny asymmetry 
between the production rates of neutrinos and antineutrinos. If the sterile 
 is lighter than the active neutrino and if the values of 
mixing parameters are in the intervals previously indicated, this initial 
asymmetry is amplified through the generation of a lepton number that starts 
at the critical temperature $T_c$.

 In this mechanism the momentum dependence plays an important role 
\cite{Foot97,DiBari99} and a monochromatic approximation provides only a 
rough description. The interesting feature is that, within this simplified 
physical picture, the quantum kinetic description through the density matrix 
collapses into an {\em effective (classic) kinetic description} 
(Pauli-Boltzmann approach). This is possible because in the regime
where collisions are rapid enough, the evolution of
the off-diagonal terms can be disentangled by the diagonal terms evolution
and explicit expressions can be derived for them.
In this way, in the description of the evolution of the statistical properties of mixed neutrinos,  only the diagonal terms are left, 
the usual statistical distributions: a simpler physical picture yields a 
simpler mathematical description. 
The derivation of the equations for the distributions can be done either 
directly from the simplified physical model \cite{Foot97} or also, 
with a more formal procedure, from the quantum kinetic equations themselves, 
via some approximations, indicated by the authors globally as 
{\em static approximation}, valid under appropriate conditions on the 
mixing parameters \cite{Bell99}. The differences that one expects from 
a full quantum kinetic description are at low temperatures 
($T\stackrel{<}{\sim}5\, {\rm MeV}$), when the MSW 
effect becomes important and must be taken into account, and at the critical
temperature for large enough vacuum mixing angles, when the growth of lepton 
number is so rapid that an adiabatic condition to describe the process
does not hold anymore. This adiabatic condition is the possibility to neglect
any change in the effective potentials, and thus in the mixing angle, 
between the collisions on average. The collisions, if vacuum 
mixing angles is not too large, are able to average out the coherent 
effects of the oscillations in the macroscopic quantity 
(like the neutrino asymmetry). 

  This analysis on the validity of the approach is confirmed by 
the numerical calculations performed using the QKE. In \cite{Foot97b,foot99}
it has been shown how the MSW effect at low temperature is able to
amplify the growth of lepton number up to values slightly lower than
the maximum absolute value obtainable of $3/8$, corresponding to a 
situation when all active antineutrinos (or neutrinos) are converted into 
sterile neutrinos and to a value $\xi_{\alpha}\simeq 0.5$ for the
chemical potential 
\footnote{
The authors of a recent paper obtain much lower values for the final 
lepton number \cite{dolgov2}. 
This result 
has been obtained
solving an approximated equation that, 
compared to the static approximation equation,  contains 
a new term responsible, according to them, 
for the different values. 
We do not hide our scepticism toward these 
results, due to the robust coherent picture 
of previous ones, obtained both in a physically
clear approximated picture \cite{Foot96, Foot97, DiBari99}
and confirmed by numerical calculations using the
exact QKE equations \cite{Foot97b,foot99}. 
The authors of \cite{dolgov2} 
try to justify this situation claiming a not rigorous 
numerical procedure in the exact numerical 
calculations, without being aware of
the conspiracy that would exist between a supposed 
numerical error in the solution of the exact QKE
and the effect of a missing term in 
the static approximation equation. 
Moreover new numerical solutions of the exact QKE 
will be soon presented together with 
clear checks of their validity \cite{DiBari99b}
and an accurate description of the adopted
numerical procedure. Although there is no final 
agreement at the present,
we wish to thank A.D. Dolgov, S.H. Hansen, S. Pastor
and D.V. Semikoz, for the kind availability to answer
our questions.
}. On the other hand
it has been shown how, for temperatures $T\stackrel{>}{\sim}5\, {\rm MeV}$,
an effective kinetic approach agrees almost perfectly with the
quantum kinetic one for small mixing angles. It is also possible
to extend the numerical study of QKE to large mixing angles
\cite{DiBari99b} 
and it has been found that for $\sin^{2} 2\theta_{0}\stackrel{>}{\sim} 10^{-6}$
and 
$\sin^{2} 2\theta_{0}\stackrel{<}{\sim}\,3\times 10^{-4}({\rm eV^{2}}/|\delta m^{2}|)$, 
in the quantum kinetic approach, at the critical temperature $T_{c}$, 
rapid oscillations in the lepton number take place. This behaviour was 
first studied in \cite{Shi96} and later confirmed in \cite{Enqvist99}
(even though in a smaller region of mixing parameters), 
but in these works the momentum dependence was not taken into account 
and the rapid oscillations are observed for a much larger region 
of mixing parameters. In \cite{DiBari99b} is concluded that, outside 
the special region of mixing parameters where rapid oscillations are 
observed at the resonance, at 
temperatures $T\stackrel{>}{\sim}5\, {\rm MeV}$,  the effective
kinetic approach provides a very 
good description not only for the evolution of the absolute value of 
lepton number but also in predicting its final sign. In this paper we 
will extend this analysis to the case when some tiny inhomogeneities 
in the baryon number (baryon isocurvature perturbations) are present, dealing 
with the region of mixing parameters where the effective kinetic approach 
can be used. This can be done including into the equation a term
that describes the diffusion of neutrinos and we will show how
this effect can induce the generation of lepton domains 
\footnote{
The possibility of a generation of lepton domains was first 
claimed in \cite{Foot96} due to a sign indetermination in the 
obtained equation for the lepton number evolution. 
 This results in a generation of lepton domains with sign randomly determined.
However in \cite{Foot97} it was shown how
the account of a correcting term produces 
a full sign determination of the solution (see Section 2).   
The idea of a "chaotic" generation of lepton domains has been 
recently re proposed in \cite{Shi99}.
  This model assumes that the lepton number undergoes at the resonance very rapid 
and unstable changes of sign for any choice of mixing parameters and this
would again result in a sign indetermination in different points of space. 
 The analysis presented in \cite{DiBari99b}, including a full momentum
dependence, excludes this possibility
for almost all values of mixing parameters, except in a special 
region where the numerical calculations cannot be, at the moment, conclusive.
In the mechanism we present here, 
the generation is not chaotic but perfectly determined by the 
spectral features of the baryon number inhomogeneities and moreover 
the horizon scale is not a limit to the size of lepton domains.
}.
 
The plan of the paper is the following.
In section 2 we will show how the problem of sign of lepton number
is fully determined in the effective kinetic approach, re analyzing 
and extending  a procedure presented in \cite{Foot97}. In section 3 we will 
extend  the analysis to the case where small 
inhomogeneities in the baryon number are present, through the introduction of 
a diffusion term. We will illustrate how this term can induce, in some regions,
an inversion in the sign of lepton number growth, bringing to the
formation of lepton domains. 
In section 4 we describe qualitatively 
the evolution of lepton domains once they have been generated. 
In section 5 we conclude and discuss the possible applications of the new 
proposed mechanism.

\section{Final sign of lepton number in a homogeneous background}

 Within the Pauli-Boltzmann approach developed in \cite{Foot96,Foot97}, 
the evolution of the lepton number carried by an active $\alpha$-neutrino flavor
that is mixed with a sterile neutrino flavor, is described by the following equation:
\begin{equation}\label{eq:lnua}
\frac{dL_{\nu_{\alpha}}}{dt}=\left[A(T,L)\mbox{ }L-B(T,L)\;L_{\nu_{\alpha}}\right]
\end{equation}
where we defined 
\footnote{
We are actually neglecting a third term that arises
only when a sterile neutrino asymmetry is produced, that means when
the lepton number of active neutrinos is changing, considering that
$L_{\nu_{s}}+L_{\nu_{\alpha}}=$const. This term can give effects only when
an initial $\alpha$-lepton number much higher than $\tilde{L}$ is assumed.
Here we do not consider this situation and in this case this third term can
be neglected.
}:
\begin{equation}\label{eq:A}
A\equiv\frac{T^{3}}{2\pi^{2}n_{\gamma}}
      \int\,dy\,y^{2}f^{0}_{eq}
(\Gamma_{\alpha s}-\bar{\Gamma}_{\alpha s})(z^{+}_{s}-z^{+}_{\alpha})
\end{equation}
\begin{equation}\label{eq:B}
B\equiv\frac{6\zeta(3)T^{3}}{\pi^{4}n_{\gamma}}
      \int\,dy\,\frac{y^{2}\,e^{y}}{(1+e^{y})^{2}}
	(\Gamma_{\alpha s}+\bar{\Gamma}_{\alpha s})
\end{equation} 
The variable $y$ in the integrals is the adimensional momentum $p/T$. 
The quantity  $L_{\nu_{\alpha}}$ is rigorously defined as the lepton 
number of the $\alpha$-active neutrino in the portion of comoving volume 
that contains a fixed number of photons $N_{\gamma}^{in}$ 
at some initial temperature 
$T_{in}\stackrel{<}{\sim}150\,\rm{MeV}$. 
Considering that the evolution of lepton number freezes at temperatures 
around $T_f\simeq  1\, \rm{MeV}$, one can safely consider the number of 
photons in the element of comoving volume as a constant 
(neglecting muon annihilations) and write:
\begin{equation}
L_{\nu_{\alpha}}\equiv 
\frac{N_{\nu_{\alpha}}-N_{\bar{\nu}_{\alpha}}}{N_{\gamma}^{in}}
\simeq\frac{n_{\nu_{\alpha}}-n_{\bar{\nu}_{\alpha}}}{n_{\gamma}}
\end{equation}
We indicated with $n$ the particle densities and with $N=nR^{3}$ 
the numbers of particles in the comoving volume $R^{3}$.
We also introduced the {\em (effective) total lepton number} 
$L$ defined as:
\begin{equation}
L \equiv L_{\nu_\alpha} + L_{\nu_e} + L_{\nu_\mu}+ 
L_{\nu_\tau} \pm {1\over 2}B_{n}  \equiv 2 L_{\nu_\alpha}+\tilde{L}
\end{equation}
where $+$ holds for $\alpha=e$ and $-$ for $\alpha=\mu,\tau$.
The field $\tilde{L}$ is the total charge number of non oscillating 
neutrinos plus a contribution from the baryon number carried by neutrons: 
we will refer to it as the {\em background charge}.
It is constant while the oscillations occur, and must be considered 
as a parameter given by some earlier phase of baryo-leptogenesis. 
In this section we will assume that it is also strictly homogeneous, 
while in the next section we will study the effect of the presence of 
inhomogeneities. Both in $L$ and in $\tilde{L}$ we dropped an index 
$\alpha$ to simplify the notation.

The quantities $\Gamma_{\alpha s}$, $\bar{\Gamma}_{\alpha s}$ are the 
production rates for sterile neutrinos and antineutrinos, $f^{0}_{eq}$ 
is the Fermi-Dirac distribution with zero chemical potential and 
$z^{\pm}_{\alpha,s}=(f_{\nu_{\alpha,s}}
\pm f_{\bar{\nu}_{\alpha,s}})/2f^{0}_{eq}$ 
are the sum and difference distributions of neutrinos, relative to the 
Fermi-Dirac one. The active neutrinos distributions 
can be safely described by thermal equilibrium distributions 
because the process of generation of lepton number occurs for
temperatures $T\gg 1 {\rm MeV}$ (the total collision rate 
is therefore much higher than the expansion rate) and for
small mixing angles (the collision rates that refill the quantum 
states of active neutrinos are much higher than the sterile 
neutrinos production rates that deplete them). On the other hand
the sterile neutrinos distributions
must be described by two other rate equations that, 
together with the expressions for the production rates, can be found 
in \cite{Foot97,DiBari99} (we do not need them in the present context). 

For a qualitative understanding of the evolution of lepton number, 
with a specific attention to its sign, it is useful to recast the 
equation (\ref{eq:lnua}) in the form:
\begin{equation}\label{eq:dlnuab}
\frac{dL_{\nu_{\alpha}}}{dt}=
  \left[2A(T,L)-B(T,L)\right]\;(L_{\nu_{\alpha}}-L_{eq})
\end{equation}
where we introduced the quantity:
\begin{equation}\label{eq:Leq}
L_{eq}=-\frac{A\,\tilde{L}}{2A-B}
\end{equation}
While the term $B$ is always positive, the term $A$ changes sign at
a critical temperature $T_{c}$, being negative for higher
temperatures and positive for lower temperatures.
When the temperature is far from the critical value
 ($|T-T_{c}|\gg \Delta T\simeq (2-3)\,{\rm MeV}$), 
the $B$ term acts as a correcting term ($B\ll |A|$) and
the fixed point $L_{eq}\simeq L^{0}_{eq}\equiv -(1/2)\,\tilde{L}$.
Moreover, until the lepton number is below a threshold value 
$L_{*}\simeq 10^{-6}(|\delta m^2|/{\rm eV^2})^{\frac{1}{3}}$ 
(see \cite{DiBari99}), its non linear effect inside $A$ and $B$ 
can be neglected. In this situation the equation (\ref{eq:lnua}) 
becomes extremely simple:
\begin{equation}
\frac{dL_{\nu_{\alpha}}}{dt}=A(T)\,L
\end{equation}
While  for temperatures above $T_{c}$  the quantity $A$ is negative, the 
fixed point $L_{eq}$ is stable and thus the total lepton 
number is destroyed, for temperatures below $T_{c}$, $A$ is positive, 
the fixed point is unstable and the total lepton number starts to grow.
To answer the question toward which direction it grows, one has to take 
into account the term $B$ and study its action in the vicinity of the 
critical temperature when $A$ is small and $B$ becomes 
important in driving the evolution of lepton number. Let us see in 
different stages what happens when temperature approaches the critical 
value from the above. When $B$ is still small compared to $A$, expanding 
up to the first order in $B/A$, one gets:
\begin{equation}\label{eq:leq}
L_{eq}=
-\frac{1}{2}\tilde{L}+\frac{1}{4}\frac{B}{|A|}\tilde{L}
\end{equation} 
During this period the term $2A-B$ is negative and the solution tracks
the fixed point $L_{eq}$. An approximate expression 
for the quantity $\delta L\equiv L_{\nu_{\alpha}}-L_{eq}$ can be
derived neglecting a term $d(\delta L)/dt$ in the Eq. (\ref{eq:dlnuab}):
\begin{equation}
\delta L=\frac{1}{2A-B}\frac{dL_{eq}}{dt}  
\end{equation}
Until $2A-B\gg dL_{eq}/dt$, the solution tracks the fixed point.
From the expression (\ref{eq:leq}) it is clear that the term $B$ drives 
the growth of the lepton number toward the same sign as the 
background charge $\tilde{L}$. At the 
critical temperature the term $A$ vanishes and simply $L_{eq}=0$.
In this moment the fixed point is changing very rapidly while 
the term $2A-B$, that should force the solution $L_{\nu_{\alpha}}$ to track 
$L_{eq}$, is small and the solution starts to diverge from the fixed point. 
Immediately below the critical temperature, when $2A=B$, the fixed point 
has a vertical asymptote and changes sign and at still lower temperatures 
it rapidly approaches zero again. This time however 
$A$ is positive and the fixed point $L_{eq}$ does not attract the
solution any more and this continues to grow toward the same 
direction transmitted by the "derailment" action of the term $B$  at the 
critical temperature. 

The behaviour of the solution around the critical 
temperature is shown in fig. 1 for a particular choice of the mixing 
parameters.

\section{Lepton domains formation in presence of inhomogeneities} 

In this section we will generalize the process of active-sterile neutrino
oscillations to the case when spatial inhomogeneities are present. 
 From the observation of CMB we know that small inhomogeneities in the 
temperature field (adiabatic perturbations) were present in the early universe. 
However taking into account the presence of these perturbations does not change 
the basic results of the homogeneous scenario. The lepton
number growth starts at slightly different times in different points,
but the final result is unchanged: the presence in the end of a final
lepton number with the same sign everywhere in the space and with the same 
absolute value as in the homogeneous case. 

More interesting is to consider the possibility that small inhomogeneities are
present in the background charge $\tilde{L}$ (isocurvature perturbations). 
This quantity is the sum of
the lepton number carried by the non oscillating neutrinos and a term 
given by the presence of a baryon number. Whatever is the mechanism that 
created the inhomogeneities, it is reasonable to think that the inhomogeneities
in the lepton numbers carried by the non oscillating neutrinos 
become soon much smoother than those in the baryon number, due to the much 
higher diffusion of neutrinos. More 
precisely we can consider the baryon number field expressed in the comoving 
coordinates as a constant, neglecting the neutron diffusion. This will make much 
simpler our following discussion, without altering the main results. 

A first trivial effect on the equation (\ref{eq:lnua}) is simply
that now $\tilde{L}=\tilde{L}({\bf x})$, where ${\bf x}$ are
the comoving coordinates.  At temperatures above $T_{c}$ this effect is
enough to understand how the process is modified by the presence of
the inhomogeneities. The $\nu_{\alpha}$-lepton number will 
evolve in a way to destroy in any point the total lepton number $L$
and a zero order solution, considering only the $A$-term, 
is given by $L_{\nu_{\alpha}}=-(1/2)\,\tilde{L}({\bf x})$.
When temperature drops down, approaching the critical value, again
the action of the correcting term $B$ must be considered in order to
understand toward which direction, positive or negative values, the total
lepton number will grow. 

This time however a new correcting term must be considered in the equation,
a term that arises due to the neutrino diffusion, and the full equation
becomes now:
\begin{equation}\label{eq:lnuad}
\frac{dL_{\nu_{\alpha}}}{dt}=A(T,L)\,\left[2L_{\nu_{\alpha}}+
                  \tilde{L}({\bf x})\right]
			 -B(T,L)\;L_{\nu_{\alpha}}
			+\frac{\cal D}{R^{2}}\nabla^{2}L_{\nu_{\alpha}}
\end{equation}
where clearly $L_{\nu_{\alpha}}=L_{\nu_{\alpha}}({\bf x},t)$, 
${\cal D}$ is the {\em diffusion coefficient} and $R$ is the scale 
factor that appears because we are dealing with comoving coordinates
(we normalize $R$ in a way that $R_{0}=1$), while
the diffusion term must be calculated in physical lengths.

 The appearance of a diffusion term is something intuitively clear.
For a more rigorous derivation one has simply to write the Liouville
operator in the left hand side of the Boltzmann equation in the 
inhomogeneous case:
\begin{equation}
 \frac{d}{dt}f_{\nu_{\alpha}}({\bf x},{\bf p},t)=
\frac{\partial}{\partial t}f_{\nu_{\alpha}}({\bf x},{\bf p},t)
-H{\bf p}\cdot{\bf\nabla}_{\bf p}\,f_{\nu_{\alpha}}({\bf x},{\bf p},t)
+\frac{c}{R}\,\hat{p}\,\cdot{\bf\nabla}_{\bf x}\,f_{\nu_{\alpha}}({\bf x},{\bf p},t)
\end{equation}
where $H$ is the expansion rate $\dot{R}/R$.
In this way, integrating on the momenta the Boltzmann equation for the 
distribution difference of active neutrinos  
$f^{-}_{\nu_{\alpha}}\equiv f_{\nu_{\alpha}}-f_{\bar{\nu}_{\alpha}}$ 
and dividing for the photon number density, one obtains:
\begin{equation}
\frac{dL_{\nu_{\alpha}}}{dt}=\left[\frac{dL_{\nu_{\alpha}}}{dt}\right]_{\rm osc}+\frac{1}{R^{2}}{\bf\nabla}_{\bf x}
\cdot\left({\cal D}\,{\bf\nabla}_{\bf x}L_{\nu_{\alpha}}\right)
\end{equation}
where the first term in the r.h. side is the contribution from the
oscillations into the sterile neutrinos (the r.h. side in the Eq. 
(\ref{eq:lnua})), while the second term is the contribution due to the 
diffusion. The diffusion coefficient is defined by the expression:
\begin{equation} 
\int \frac{d^{3}p}{(2\pi^{3})} f^{-}_{\nu_{\alpha}}({\bf x},{\bf p},t)\,\hat{p}
\equiv  -\,\frac{\cal D}{R}\,{\bf\nabla}_{\bf x}
\int\frac{d^{3}p}{(2\pi^{3})} f^{-}_{\nu_{\alpha}}({\bf x},{\bf p},t)
\end{equation}
As we are considering the case of small inhomogeneities in the baryon number,
the diffusion coefficient can be safely considered homogeneous and
one gets the diffusion term in the form written in the equation (\ref{eq:lnuad}).

The order of magnitude of ${\cal D}$ is given by 
$c\,{\ell}_{int}$, where ${\ell}_{int}=\Gamma_{int}^{-1}$ is the interaction
length and $\Gamma_{int}=3.15\,k_{\alpha}G^{2}_{F}T^{5}$ is the total 
collision rate, with $k_{\alpha}\simeq 1.27\mbox{ }(0.92)$ \cite{Enqvist92b} 
for $\alpha=e\,(\mu,\tau)$ 
\footnote{An explicit calculation of the diffusion coefficient in the 
simpler case of "light" particles diffusing in a 
"heavy" particles medium  can be found in \cite{landau}.
}.

 If the diffusion term is negligible compared to the $B$ term, then the
homogeneous scenario is practically unchanged. In the equation (\ref{eq:leq})
one has simply to consider that now $\tilde{L}=\tilde{L}({\bf x})$ and, as 
we are considering small inhomogeneities so that the sign of $\tilde{L}$ is
spatially constant, the $B$ term pushes the solution toward the same direction
in all the points. 

On the other hand if we assume that the diffusion term is dominant, the 
situation can be much different. Let us assume for definiteness that
$\tilde{L}>0$. In the regions where the background charge 
is lower, the $\alpha$-neutrino lepton number, before the critical 
temperature, would be higher (because $L_{\nu_{\alpha}}\simeq -0.5 \tilde{L}$). 
In these regions the diffusion term pushes the lepton number to be depleted 
with the result that at the critical temperature one can have 
$L_{\nu_{\alpha}}< -0.5 \tilde{L}$ and the growth  
starts toward a negative sign. The vice versa would happen in the 
regions where the background charge is higher. The final result is the 
creation of regions with different sign of lepton number with the 
same size as the scale of baryon number inhomogeneities. The sign of 
lepton number is in fact determined at any point by the curvature of 
the background charge field and thus the global properties of the
baryon number inhomogeneities are somehow transmitted to the lepton domains. 

We have now to study the conditions for which this mechanism can work. 
The involved parameters are the mixing parameters, the amplitude of the 
inhomogeneities and the size of the inhomogeneities. We can assume, for 
simplicity, that inhomogeneities have only one size scale. Let us 
introduce the field $\rho({\bf x})$ of the inhomogeneities, writing 
the {\em background charge} field as:
\begin{equation}
\tilde{L}({\bf x})=\bar{L}\left[1+\rho({\bf x})\right]
\end{equation}
where $\bar{L}$ is the mean value, and where we assume that the 
field $\rho$ is a perturbation ($\rho\ll 1$ at any point).
As we already said, while also non oscillating neutrinos give a non 
negligible contribution to the mean value, the baryon number 
gives the dominant contribution to the inhomogeneities. 

At temperatures higher than $T_{c}$ the solution will approach
again the fixed point that at the zero order will be simply
$L_{eq}^{0}({\bf x})=-(1/2)\tilde{L}({\bf x})$. The final sign of lepton 
number will be again determined by the correcting terms that this time 
include also the diffusion term. This can be generically estimated 
through the following expression (we drop the
dependence on ${\bf x}$ in $\rho$ and $L^{0}_{eq}$):
\begin{equation}\label{eq:D} 
\frac{\cal D}{R^{2}}\nabla^{2}L_{\nu_{\alpha}}
\simeq  -\Gamma_{int}\,\rho
\left(\frac{\ell_{int}^{(0)}}{\lambda^{(0)}}\right)^{2}\,L_{eq}^{0}
\equiv -D\,L_{eq}^{0}
\end{equation}
where we indicated with $\lambda$ the scale of the inhomogeneities
and, given the generic physical length ${\ell}$, we mean
with ${\ell}^{(0)}=(R_{0}/R)\,{\ell}$, the comoving length 
normalized at the present. We also introduced the 
convenient quantity $D$.
In this way the expression (\ref{eq:Leq}), valid for the fixed point 
in the homogeneous case, becomes:
\begin{equation}
L_{eq}=-\frac{A\,\tilde{L}}{2A-B-D}
\end{equation}
or at the first order in $(B+D)/A$: 
\begin{equation}
L_{eq}=-\frac{1}{2}\tilde{L}+\frac{1}{4}\frac{B+D}{|A|}\tilde{L}
\end{equation}
This expression is now describing what we previously said in words:
if the term $D$ is negative (it means $\rho<0$, the regions
where the background charge is lower than the mean value if we assume 
it is positive) and its absolute
value higher than $B$, then the lepton number growth is addressed toward
the opposite sign than in the regions where the opposite condition holds.

To determine the conditions for which an inversion of sign is possible,
we have thus simply to compare the term $B$ with the absolute value of $D$
at temperatures around the critical temperature.
An estimation of  the order of magnitude of $B$ can be easily done considering that
the integral on the momenta receives a dominant contribution only
around the resonant value. This procedure has already been employed
in \cite{DiBari99} (in that paper it was used to estimate the $A$-term)
and here we only give the result (valid for temperatures 
$T\stackrel{>}{\sim} T_{c}$, 
when the lepton number is still small and can be neglected):
\begin{equation}
B\sim 10^{2}\,s^{2}\;\Gamma_{int}
\end{equation}
where we defined $s\equiv \sin 2\theta_{0}$. 
A calculation of the comoving interaction length expressed in parsec 
gives the result:
\begin{equation}
{\ell}^{(0)}_{int}\simeq 100\,(70)\,{\rm pc}\,\left(\frac{\rm MeV}{T}\right)^{4}
\end{equation}
for $\alpha=\mu, \tau (e)$.
Using this expression and imposing the condition $|D|>B$ for the generation
of lepton domains, it is thus straightforward to derive the following 
condition on the (orders of magnitude) of the involved parameters:
\begin{equation}\label{eq:rho}
\rho\left(\frac{\lambda^{(0)}}{pc}\right)^{-2}\stackrel{>}{\sim} 
     10^{7}\,s^{2}\,\left(\frac{|\delta m^{2}|}{\rm eV^{2}}\right)^{\frac{4}{3}}
\end{equation}

This analytical expression is confirmed by the numerical calculations.
In figure 2 we show the evolution of the terms $|A|$, $B$,
calculated numerically for a particular choice of the mixing 
parameters and compared with the term $|D|$, calculated 
through the expression (\ref{eq:D}) with different values of 
the parameter $|\rho|/(\lambda^{(0)})^{2}$. The result
is in agreement with the one that could be derived using 
the eq. (\ref{eq:rho}).

This condition can also be expressed as a condition on the order of magnitude
of the maximum size of lepton domains that can be generated 
\footnote{
There is also a condition on the minimum size
of lepton domains that can be generated \cite{footpc}:
$\lambda^{(0)}\stackrel{>}{\sim}{\ell}^{(0)}_{int}(T\simeq T_{c})
\sim 10^{-3}\,{\rm pc}(|\delta m^{2}|/{\rm eV^{2}})^{-{2\over 3}}$.
Below this scale neutrino are free streaming and they
can destroy the lepton domains more rapidly than they can be generated,
considering that $\dot{L}\stackrel{<}{\sim}\Gamma_{int}$.
}:
\begin{equation}
\lambda^{(0)}\stackrel{<}{\sim}
10\,{\rm Kpc}
	\left[\frac{|\delta m^{2}|}{\rm 10^{-5}\,eV^{2}}\right]^{-\frac{2}{3}}
	\left(\frac{\rho}{10^{-1}}\right)^{\frac{1}{2}}\,
	\left(\frac{10^{-10}}{s^{2}}\right)^{\frac{1}{2}}
	\stackrel{<}{\sim}10\,{\rm Kpc}
\end{equation}
In this last expression we indicated the extreme possible values of 
the different quantities that yield the maximum value for $\lambda^{(0)}$. 
In particular the condition $s^{2}\stackrel{>}{\sim}10^{-10}$ must be 
imposed to have a not negligible final absolute value of lepton number.

In next section we will use this result to sketch a scenario 
of how lepton domains with different sizes evolve and 
affect the BBN. This analysis will suggest a way to 
circumvent the upper limit found on the 
maximum size of a lepton domain.

In conclusion of this section we notice that because of the presence
of the term $B$ there are always points where, even though $\rho<0$, the 
condition $B-|D|<0$ is not verified. 
Therefore we can say that the $B$
term will always favour a dominance of regions where the 
generation of lepton number occurs with the same sign of $\tilde{L}$:
we will refer to it as the {\em dominant sign}, compared to 
regions with {\em inverted sign}.
There are however two much different extreme situations of dominance.
In a first one $B/|D|\ll 1$ everywhere, 
except in thin walls
around the surfaces where $\rho=0$ ({\em weak dominance}). 
In this case the size
of lepton domains coincides with the size of baryon number inhomogeneities
and in a comoving sphere with a much larger radius than the size of
inhomogeneities,  the volume of regions with dominant sign lepton number 
is almost equal to that of regions 
with inverted sign lepton number. The topology
of lepton domains is that one of a cubic lattice with each cube 
surrounded by first closest neighbours cubes with opposite sign and 
second closest neighbours cubes with the same sign.

In a second extreme situation the condition $B-|D|<0$ is verified
only in small regions around the points where $|\rho|$ is maximum,
with a size much smaller than that of the inhomogeneities
({\em strong dominance}). 
In this case one has a  structure of lepton domains with 
inverted sign (3-dim) islands  in a dominant sign background. 

\section{Lepton domains evolution}

Lepton domains start to be formed at $T=T_{c}$. With the increase of lepton number
the diffusion at the border of domains also increase. There is some 
interplay between the rate of generation (the $A$-term) and the rate of 
diffusion. 

Let us assume that the dominant sign is positive.
In this case the diffusion can be described as a process that gradually
fills at the border the regions with negative sign (the "holes").
At the same time, in the positive sign regions, the lepton number that 
has been used to fill the holes, is restored by the oscillations through
the growth term. In this situation it is clear that if the diffusion is able
to cover the whole hole, the generation of lepton domains is only
a transient regime without any practical effect.
We can refer to this phase as {\em cannibalization regime}, in which the
dominant sign regions enlarge at the expense of the inverted sign regions, until
eventually their complete disappearance.

However this process can last only until the generation of a lepton number 
can occur at the border of domains, where there is a change of sign and the
lepton number is kept small by the diffusion. When the
temperature drops down to $T\simeq 0.6\,T_{c}\,$ 
\footnote{
It corresponds to a situation when the resonant neutrinos 
have a momentum $p\stackrel{>}{\sim}10\,T$, in the tail 
of the distribution. For small values of $L$,
neutrinos and antineutrinos are both resonant but if
$L>0\,(<0)$ and if $y\geq y_{\rm peak}\simeq 2.2$, the number of 
resonant antineutrinos (neutrinos) is a little bit higher than
the number of resonant neutrinos (antineutrinos): this explains why
the fixed point $L\simeq 0$ is unstable and lepton number starts to grow
(see \cite{DiBari99}).}  
or in any case down to $3$ {\rm MeV},
the growth starts to be inhibited.  
We have to calculate the diffusion length to know how large the
lepton domains with inverted sign must be in order to survive
to the cannibalization regime. 
This is given by the following expression that takes into account the 
Universe expansion:
\begin{equation}
\lambda_{d}^{(0)}=\int_{t_{c}}^{t}\frac{v_{d}\,dt'}{R(t')}
\end{equation}
The {\em diffusion velocity} is approximately given by 
$v_{d}\sim \sqrt{{\cal D}/(t-t_{c})}$. Considering that $t\gg t_{c}$ and
that $t\sim {\ell}_{H}$, it is an easy task to get the following 
expression for the comoving diffusion length:
\begin{equation}
\lambda_{d}^{(0)}\simeq 100 (70)\,{\rm pc}
\left[\left(\frac{\rm MeV}{T}\right)^{\frac{5}{2}}
				-\left(\frac{\rm MeV}{T_{c}}\right)^{\frac{5}{2}}\right],
\end{equation}

for $\alpha=\mu,\tau\,(e)$.
The generation of lepton number can only occur down to temperatures
$T\simeq \min(3\,{\rm MeV},0.6\,T_{c})$, therefore for 
$T_{c}\stackrel{>}{\sim}5\,{\rm MeV}$,
using the expression (\ref{eq:TC}), we get:
\begin{equation}\label{eq:min}
\lambda_{d}^{(0)}(T=0.6 T_{c})\sim 0.1{\rm pc}\,
		\left(\frac{|\delta m^{2}|}{\rm eV^{2}}\right)^{-\frac{5}{12}}
\end{equation}
For sizes below this value ({\em very small scales}) lepton domains with 
inverted sign are destroyed before they can produce any effect.
For $T_{c}\stackrel{<}{\sim}5\,{\rm MeV}$ the diffusion length 
between the beginning of the generation of lepton number at $T_{c}$ 
and its end at $T\simeq 3{\rm MeV}$ tends to zero for 
$|\delta m^{2}|\rightarrow 10^{-5} {\rm eV^{2}}$, 
but in any case there is a lower limit on the size of lepton domains
that can be generated \cite{footpc} and in the end the same expression
(\ref{eq:min}) can be approximately used.

  For larger scales  the cannibalization cannot destroy completely the 
lepton domains before that the generation of lepton number at the border 
stops. When this happens, the lepton number that diffuses to fill the hole 
is not generated any more: the result is an effect of dilution of 
lepton number. The generation of lepton domains leaves a mark 
that modifies the homogeneous scenario.

In this case, to understand the fate of lepton domains, it is useful
to compare the scale of lepton domains with the 
horizon scale, given by the following expression:
\begin{equation}
{\ell}^{(0)}_{H}=100{\rm pc}\,\left(\frac{\rm MeV}{T}\right)
\end{equation}
It coincides with the diffusion length at $1\rm {MeV}$ when neutrinos
start to free stream. All lepton domains with {\em small scales} 
$\lambda^{(0)}\stackrel{<}{\sim}{\rm 100 pc}$, do not survive and one 
gets in the end a homogeneous lepton number field.
However, as already stated, this time the generation of lepton domains  
produced the effect 
to dilute the final value of lepton number. If the condition for the 
formation of domains (\ref{eq:rho}) is satisfied only in the peaks of 
the inhomogeneities ({\em strong dominance}), then the contribution 
from the regions with inverted sign 
is negligible and the dilution effect too.
Otherwise in the other extreme case, when the relation (\ref{eq:rho}) is 
satisfied in almost all the space ({\em weak dominance}), then 
there would be an
almost total reciprocal cancelation of regions with opposite sign, with only 
a small relic value of lepton number due to the presence of the $B$-term. All  
intermediate values of lepton number are possible.

Scales of inhomogeneities $\lambda^{(0)}\stackrel{>}{\sim}{\rm 100 pc}$ 
can produce lepton domains able to survive until the freezing of
neutron to proton ratio and therefore the presence of these
scales would give rise to an inhomogeneous BBN scenario. 

 Another important distinction is between scales able to 
produce primordial nuclear abundance inhomogeneities that survive
until the present ({\em large scales}) or not ({\em medium scales}).
In fact, even though nuclear abundances are produced in a inhomogeneous 
scenario of BBN, subsequent astrophysical mixing mechanisms, such as 
shock waves induced by supernovae explosions, would be able
to homogenize the products of BBN \cite{Jedamzik95}. These processes 
are effective on scales $\lambda^{(0)}\stackrel{<}{\sim}100 {\rm Kpc}$. 
From this point of view the mechanism we proposed, in its simplest 
version, is unable to produce visible primordial abundance
inhomogeneities.  
 
 We can however circumvent this limit if we relax our initial assumption 
on the presence of only one characteristic scale in the field
of the inhomogeneities $\rho({\bf x})$. We can in fact imagine a 
simple extension with two characteristic scales contemporarily present 
(for examples two Fourier modes): one small scale and one large scale. 
The amplitude of the first one could be modulated by the presence 
of the second. In this case one can have the simultaneous existence
of regions where the small scales inhomogeneities are able to efficiently 
dilute the lepton number generated at the critical temperature down to 
negligible values, and regions where the amplitude of small inhomogeneities 
is depressed and there is no dilution effect. In this way one can produce 
islands where lepton number is present in a background with zero lepton 
number or vice versa. This time the size of the island regions has no 
intrinsic limit, but it is a feature of the model that
produced the baryon inhomogeneities.

More realistically one has to think that the inhomogeneities 
are described by a spectrum of lengths.   
 The interesting property is that the spectral features 
of the isocurvature perturbations would be reflected in the lepton 
domains features and eventually in the BBN products. Another interesting 
aspect is that very large scale lepton domains would not be constrained 
by microwave background observations as models where large amplitude 
isocurvature perturbations are directly present in the baryon number 
\cite{Copi96}. This because the inhomogeneity in the active neutrino 
energy density would be compensated by an opposite inhomogeneity 
in the sterile neutrino energy density. In fact during active-sterile 
neutrino oscillations one has that $L_{\nu_{s}}+L_{\nu_{\alpha}}=$const.  
Even though sterile neutrinos free stream while active neutrinos are
still diffusing, for very large scales (surely larger than the diffusion
length of active neutrinos) this different behaviour cannot change
the conservation of the sum of the asymmetries
of active neutrinos and sterile neutrinos at each point. 

In the end of this section we
want also to show that small amplitudes of inhomogeneities
are sufficient for the mechanism to work.
From the expression (\ref{eq:rho}), imposing the lower limit (\ref{eq:min}) 
on $\lambda^{(0)}$ 
(in order to have lepton domains with inverted sign large enough not 
to be cannibalized), one gets the following condition on the mixing parameters:
\begin{equation}\label{eq:domains}
s^{2}\,\left(\frac{|\delta m^{2}|}{\rm eV^{2}}\right)^{\frac{1}{2}}
< 10^{-5} \,\rho 
\end{equation}
that is satisfied for large regions of possible values
of the mixing parameters, even for very small values of $\rho$.
These regions are represented in figure 3 for three different values of $\rho$.
It can be noticed that even for $\rho=10^{-5}$, there is a 
significant region where lepton domains can be generated and 
produce interesting effects. 

\section{Conclusions}

We proposed a mechanism according to which active-sterile 
neutrino oscillations would amplify small baryon number inhomogeneities 
generating a structure of lepton domains. The assumption on the
presence of baryon number inhomogeneities is a very reasonable one, 
because very small amplitudes are sufficient to produce lepton domains 
large enough to give some effect. The structure of the lepton domains 
and its effect on the BBN would reflect the spectral features of the 
same perturbations that seeded the formation of the domains. The main 
attraction of this mechanism is that the size of these domains is not 
limited by the horizon scale.  Moreover large scale lepton domains 
cannot be ruled out by current CMB observations (but maybe could be
constrained or discovered in future experiments). 
 
 The mechanism presents different applications. It provides a scenario
for a non standard BBN  (see \cite{Jedamzik98} for a review and see
also the recent paper \cite{Whitmire} where the effects of
inhomogeneous chemical potentials on BBN are considered). 
If the active neutrino is an electron neutrino the effects on the 
nuclear abundances can be remarkable \cite{Robert,shi99b} and some of them could help
to explain the present observational picture, for example 
claimed inhomogeneities in the D/H abundances from 
measurements in quasar absorption systems \cite{Webb97}. 
From this point of view one possibility is that we live in
a background of zero electron lepton number but some observed
high redshift absorption systems, for example the one observed with an 
high abundance, could be included in an island of non zero lepton number
\footnote{
The possibility that neutrino chemical potentials
inhomogeneities can be responsible for the primordial
deuterium abundance inhomogeneities able to explain the
observations from high redshift quasar absorbers,
has been proposed in \cite{dopa}.
A recent analysis of the current measurements 
of primordial deuterium abundance in high redshift
quasars absorbers can be found in \cite{kirkman99}.
}.

  We also mention that isocurvature perturbations can have interesting 
effects on large scale structure \cite{peebles87,Cen93}. 

Very generically we can say that the generation of lepton
domains enlarge the possibility of the existence of some observational 
signature of active-sterile neutrino oscillations in the early universe.
In the very fortunate case that active-sterile neutrino oscillations,
with the right parameters for a lepton domain formation, 
do really occur in nature, then we would have a powerful
probe for baryogenesis models that predict a spectrum
of inhomogeneities. The simple non observation of a signature
of the generation of lepton domains would put strong constraints. 
We conclude saying that the planned earth
experiments will be able in next future to test the hypothesis
of active-sterile neutrino oscillations and thus to rule out or 
support the proposed mechanism.

\vspace{10mm}

{\bf Acknowledgements}

\vspace{5mm}

P. Di Bari acknowledges many interesting and stimulating discussions 
with R. Foot, P. Lipari, M. Lusignoli and R.R. Volkas
and is grateful to them for many suggestions 
that made this manuscript clearer. He also
thanks K. Jedamzik for a nice discussion 
during COSMO-99.

\newpage
\section*{Figure Captions}

\hspace{8mm}{\bf Fig. 1} The behaviour of $L_{\nu_{\alpha}}-L_{eq}^{0}$ 
		 (solid line) around the critical 
		 temperature when the sign is determined by the action of the 
		 correcting term $B$. The dotted line is the quantity
		 $L_{eq}-L_{eq}^{0}$, with $L_{eq}$ 
		  defined by the expression (\ref{eq:Leq}) and
		  $L_{eq}^{0}=-(1/2)\tilde{L}$. We used
		 $\tilde{L}=10^{-10}$, while the values of mixing parameters
		 are those indicated.\vspace{5mm}

{\bf Fig. 2} Comparison of the quantities $A, B, D$ times $|dt/dT|=1/HT$ 
             (for numerical calculations it is more convenient to replace time with
		  temperature). The term $A$
		 vanishes at the critical temperature $T\simeq 15 \rm{MeV}$. The 
		 values of mixing parameters are those indicated. The diffusion term
		 is plotted for different values of the parameter 
		  $\rho/(\lambda^{(0)}\,{\rm pc}^{-1})^{2}$. 
		  It can be noticed that the diffusion term $|D|$ is dominant on 
		 $B$, around the critical temperature, for 
		 $\rho/(\lambda^{(0)}\,{\rm pc}^{-1})^{2}\stackrel{>}{\sim}10^{-1}$, 
		 as it could be deduced by the expression (\ref{eq:rho}).\vspace{5mm}

{\bf Fig. 3} Represantation of the regions where the condition (\ref{eq:domains}) 
             for the surviving of lepton domains to the canibalization 
		 regime is satisfied. The regions are those below the three 
		 solid lines and correspond to the three indicated values 
		 for the amplitude $\rho$ of the baryon number inhomogeneities that 
		 seed the lepton domain formation. Above the dotted line the 
		 contribution of sterile neutrinos to the number of effective 
		 neutrinos during the BBN is higher than $0.8$ for $\alpha=e$ 
		 (see \cite{Enqvist92b}). Even for
	       very small amplitudes ($\rho=10^{-5}$) there is a significant
		 region of parameters where the condition (\ref{eq:domains})
		 is satisfied.

\newpage


\begin{figure}[hbt]
\psfig{file=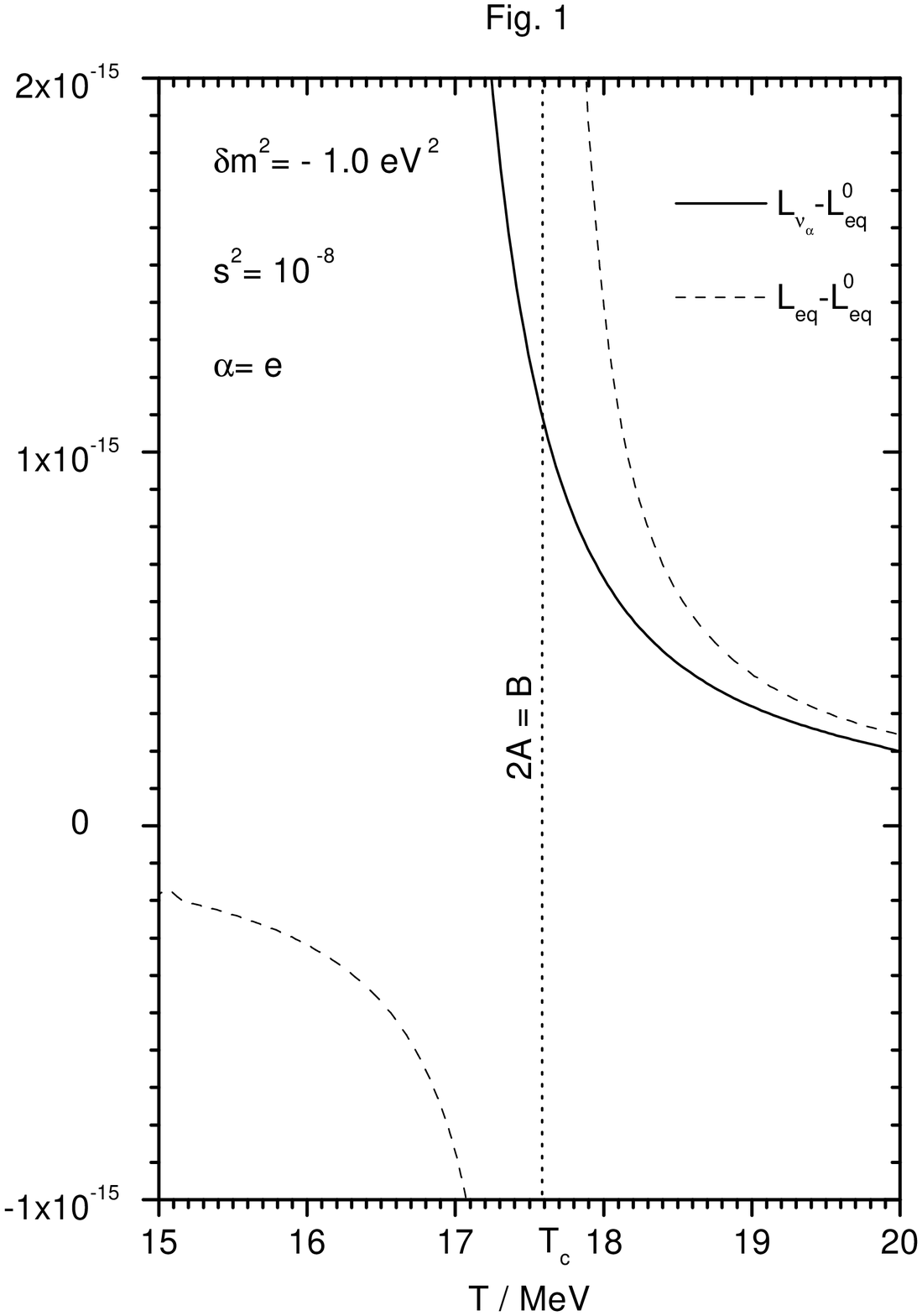,width=140mm,height=180mm}
\end{figure}

\newpage 


\begin{figure}[hbt]
\psfig{file=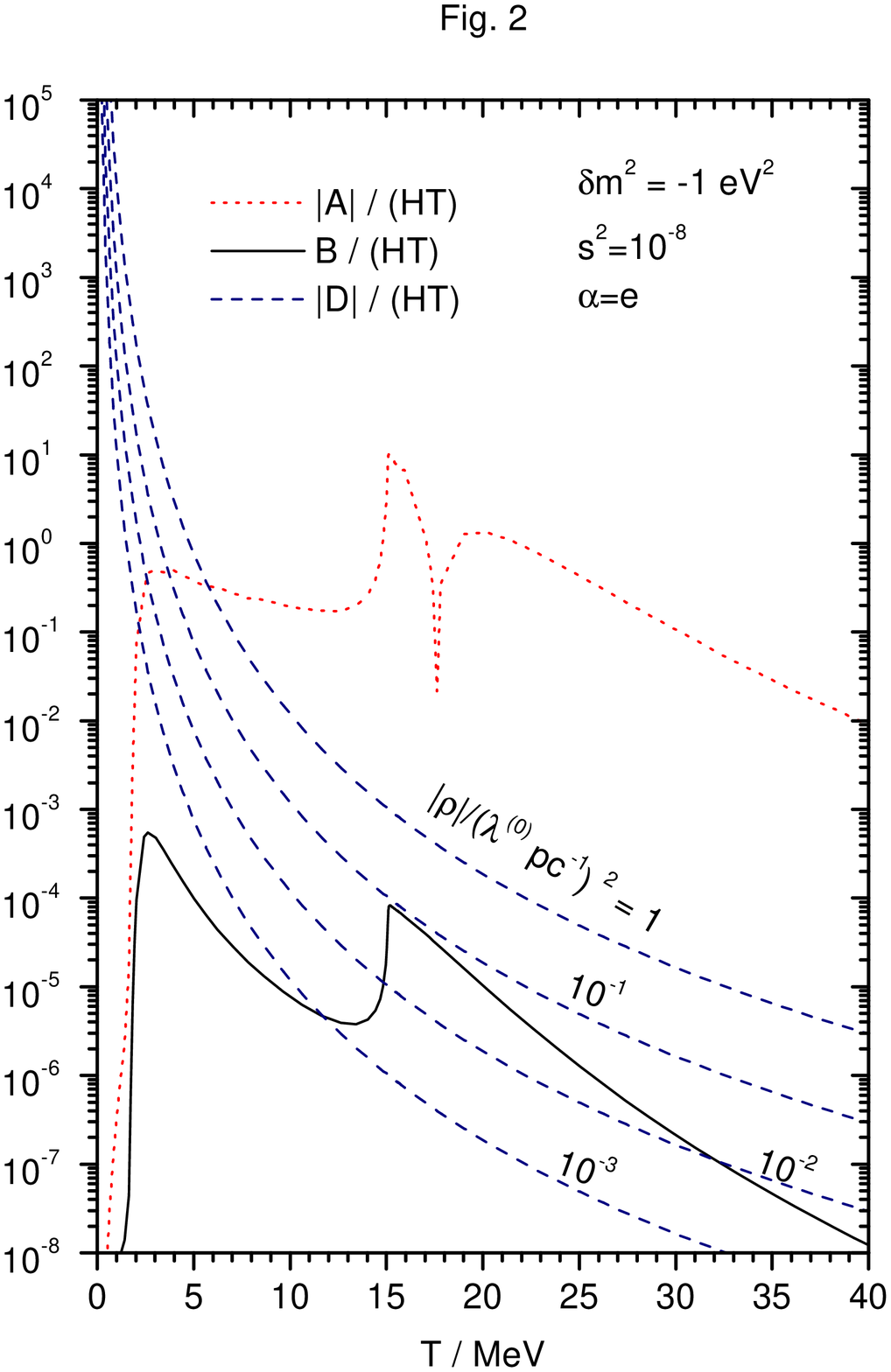,width=140mm,height=180mm}
\end{figure}

\newpage

\begin{figure}[hbt]
\psfig{file=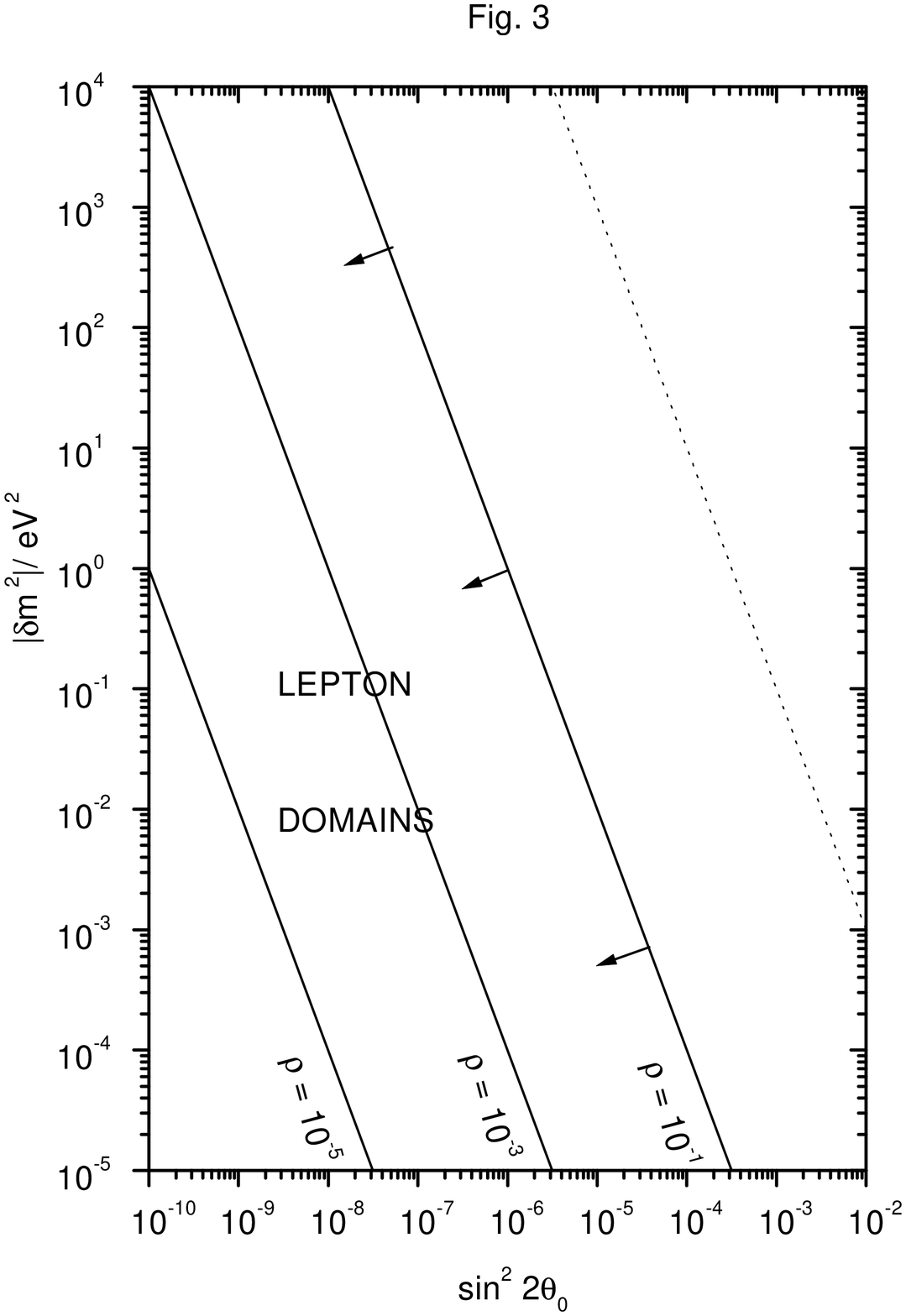,width=140mm,height=180mm}
\end{figure}

\end{document}